\documentclass[aps,prl,reprint,superscriptaddress,nofootinbib,amsfonts,amssymb,amsmath]{revtex4-1}
\usepackage[bbgreekl]{mathbbol}

\usepackage{amsmath,amssymb}
\usepackage{comment}
\usepackage[dvipdfmx]{graphicx}
\usepackage{graphicx}
\usepackage{color}
\usepackage{braket}
\usepackage{slashed}
\usepackage[bookmarks=true,bookmarksnumbered=true,setpagesize=false]{hyperref}
\usepackage{enumerate}
\usepackage{booktabs}
\DeclareMathOperator{\ephogehoge}{\epsilon}
\newcommand{\ep}[1]{\sideset{}{\scriptstyle{[#1]}}\ephogehoge}

\DeclareMathOperator{\tr}{tr}

\DeclareMathOperator{\diag}{diag}

\newcommand{\commutator}[2]{\left[#1,\,#2\right]}

\newcommand{\al}[1]{\begin{align}#1\end{align}}
\newcommand{\als}[1]{\begin{align*}#1\end{align*}}
\newcommand{\ov}{\over}
\newcommand{\nn}{\nonumber\\}
\newcommand{\tx}{\text}

\newcommand{\paren}[1]{\left(#1\right)}
\newcommand{\sqbr}[1]{\left[#1\right]}
\newcommand{\ab}[1]{\left|#1\right|}

\newcommand{\fn}[1]{\!\left(#1\right)}
\newcommand{\pa}[1]{\left(#1\right)\!{}}

\newcommand{\Paren}[1]{\bigl(#1\bigr)}

\newcommand{\Fn}[1]{\!\bigl(#1\bigr)}

\newcommand{\bs}{\boldsymbol}

\newcommand{\df}{\text{d}}

\newcommand{\mc}{\mathcal}

\newcommand{\p}{\partial}

\newcommand{\pr}{\prime}

\newcommand{\MP}{M_\text{P}}

\newcommand{\ba}{\textbf{a}}
\newcommand{\bb}{\textbf{b}}
\newcommand{\bc}{\textbf{c}}
\newcommand{\bd}{\textbf{d}}
\newcommand{\be}{\textbf{e}}

\newcommand{\diff}{{\sf diff} }

\newcommand{\B}{\text{B}}

\begin{document}
\begin{flushright}
OU-HET-1047
\end{flushright}
\vspace{-0.5cm}
\title{Dynamically emergent gravity from hidden local Lorentz symmetry}

\author{Shinya \surname{Matsuzaki}}
%\email{synya@jlu.edu.cn}
\affiliation{Center for Theoretical Physics and College of Physics, Jilin University, Changchun 130012, China}

\author{Shota \surname{Miyawaki}}
%\email{miyawaki.dl@het.phys.sci.osaka-u.ac.jp}
\affiliation{Department of Physics, Osaka University, Osaka 560-0043, Japan}

\author{Kin-ya \surname{Oda}}
%\email{odakin@phys.phys.sci.osaka-u.ac.jp}
\affiliation{Department of Physics, Osaka University, Osaka 560-0043, Japan}

\author{Masatoshi \surname{Yamada}}
%\email{m.yamada@thphys.uni-heidelberg.de}
\affiliation{Institut f\"ur Theoretische Physik, Universit\"at Heidelberg, Philosophenweg 16, 69120 Heidelberg, Germany}

\begin{abstract}
Gravity can be regarded as a consequence of local Lorentz (LL) symmetry, which is essential in defining a spinor field in curved spacetime. 
The gravitational action may admit a zero-field limit of the metric and vierbein at a certain ultraviolet cutoff scale such that the action becomes a linear realization of the LL symmetry. Consequently, only three types of term are allowed in the four-dimensional gravitational action at the cutoff scale: a cosmological constant, a linear term of the LL field strength, and spinor kinetic terms, whose coefficients are in general arbitrary functions of LL and diffeomorphism invariants. In particular, all the kinetic terms are prohibited except for spinor fields, and hence the other fields are auxiliary. Their kinetic terms, including those of the LL gauge field and the vierbein, are induced by spinor loops simultaneously with the LL gauge field mass. The LL symmetry is necessarily  broken spontaneously and hence is nothing but a hidden local symmetry, from which gravity is emergent.
\end{abstract}

\maketitle

\section{Introduction and summary}
Quantization of gravity has been one of the most profound problems in physics for more than a century.
It is known that the conventional metric theory starting from the Einstein-Hilbert (EH) action is perturbatively non-renormalizable and requires infinite amount of free parameters in the counter terms.

What degrees of freedom  (DOF) should be used in the formulation of quantum gravity?
First, the metric DOF is not sufficient and the vierbein is necessary even for just defining a spinor field in curved spacetime~\cite{Geroch:1968zm}, namely a matter field in our universe.\footnote{
One may trade and get rid of the vierbein degrees of freedom by rather promoting the gamma matrix $\gamma_\mu\!\left(x\right)=e^{\mathbf a}{}_\mu\!\left(x\right)\gamma_{\mathbf a}$ as a dynamical matrix variable; the fluctuation of $\gamma_\mu\!\left(x\right)$ can be decomposed into that of metric and $SL(4,\mathbb C)$ transformation; if one assumes that this $SL(4,\mathbb C)$ transformation is not anomalous, one may get rid of it from the path integral by assumption (barring higher dimensional operators that include derivatives of $\gamma_\mu\fn{x}$ in the action)~\cite{Gies:2013noa}. We leave this issue open in this Letter, and choose to take the vierbein as the fundamental degrees of freedom. %(In Ref.~\cite{Gies:2015cka}, it has been shown that by regarding $\gamma^\mu\fn{x}$ as a field, we may write its field configuration on a two-sphere without having a point of vanishing $\gamma^\mu\fn{x}$ as a whole. This is trivially so because $\gamma^\mu\fn{x}$ is a matrix and hence one may arrange in such a way that, everywhere, only a part of its component becomes zero. There is nothing wrong having two charts including the north and south poles, respectively, and writing a non-vanishing field on each of them, while putting a non-trivial transition function between the fields on these charts around the equator, as in the vielbein formalism; anyway this kind of charting is necessary to define a general manifold at all.)
}
In this sense, the vierbein formalism is more fundamental than the metric formalism. In paving the way to quantum gravity, we thus discard the standard folklore of the metric theory and look at the more fundamental vierbein DOF that composes the metric.
When we place a spinor field in a curved spacetime, the local-Lorentz (LL) symmetry is indispensable.
Indeed gravity can be regarded as a LL gauge theory as argued in~\cite{Utiyama:1956sy,Kibble:1961ba,Sciama:1964wt}.
Moreover, in formulating quantum gravity, it is important to identify what the path-integrated off-shell DOF of the theory is.
Regarding this point, the existence of the LL gauge symmetry in the vierbein formalism naturally leads to the idea that the LL gauge field  is also a dynamical DOF.
These facts motivate us to formulate quantum gravity in terms of the vierbein $e$ and the LL gauge field $\omega$ as independent DOF.\footnote{
We comment that the distinction whether we take $\omega$ as independent DOF leads to a physically observable difference when we start from an action with an inflaton-dependent conformal factor $\int \Omega\fn{\phi}e\wedge e\wedge F$ such as in Higgs inflation (which is one of the best-fit model so far)~\cite{Bauer:2008zj}; see also Ref.~\cite{Jinno:2019und} for a possible issue.
}

We highlight here that there is a distinct difference between the LL and ordinary Yang-Mills (YM) gauge theories. In the latter, the lowest-order gauge-invariant action starts from the kinetic term $\int\tr\fn{G\wedge\star G}$ that is quadratic in the field strength $G=\df A+A\wedge A$ of the YM gauge field $A$. On the other hand, the lowest order action for $\omega$ starts from a linear action $\int e\wedge e\wedge F$ in its field strength $F=\df\omega+\omega\wedge\omega$.\footnote{
Here the contraction of LL indices is understood as $\epsilon_{\ba\bb\bc\bd}\int e^\ba\wedge e^\bb\wedge F^{\bc\bd}$.
}
The crucial difference arises due to the existence of the vierbein $e$, which is a vector field that transforms as a fundamental representation of the LL symmetry.
We emphasize that, at this level, both $e$ and $\omega$ do not have a kinetic term and then are regarded as auxiliary fields.

At classical level, solving the equation of motion for $\omega$ requires an introduction of the {\it inverse} vierbein $e^{-1}$. 
Naively substituting the solution for $\omega$, the gauge invariant linear action $\int e\wedge e\wedge F$ turns into the EH one, which contains the kinetic term of the vierbein.
At quantum level, however, the fluctuation of $e^{-1}$ contains an infinite number of terms with unlimitedly higher powers of the fluctuation of $e$, and this theory after integrating out $\omega$ necessarily becomes perturbatively non-renormalizable.

In this Letter, we propose that the inverse, $e^{-1}$, does not exist in the bare action at a certain ultraviolet (UV) cutoff scale $\Lambda$ and that $e^{-1}$ is induced at quantum level by spinor fluctuations at lower scales.
In particular, the LL gauge kinetic term $\int F\wedge\star F$ contains $e^{-1}$ in $\star F$, so that it is prohibited at $\Lambda$.
%If it is induced by the spinor loop, 
If the $\omega$ kinetic term is induced by the spinor loop, as well as the mass term, 
the dynamical generation of this massive vector boson can be understood 
as a spontaneous breaking of the LL gauge symmetry, where the longitudinal mode 
of $\omega$ eats a part of $e$.
%this mechanism can be understood as a spontaneous breaking of the LL gauge symmetry:
The LL gauge field $\omega$ is an auxiliary field at $\Lambda$, whereas it becomes dynamical and, at the same time, massive at lower scales.

Indeed, such a phenomenon is observed in quantum chromodynamics (QCD): 
There, the $\rho$ meson can be understood as a gauge boson for the so-called hidden local gauge symmetry;
see e.g.\ Ref.~\cite{Bando:1987br} for a review. 
Namely, we claim that the LL gauge symmetry is a hidden local symmetry.
After integrating out $\omega$ at low energy, the EH term made of both $e$ and $e^{-1}$ is an effective operator as a consequence of the symmetry breaking.\footnote{
Throughout this Letter, $\omega$ denotes the LL gauge field and should not be confused with the $\omega$ meson in QCD.
}

How can we guarantee the above scenario?
In particular, can we naturally prohibit $e^{-1}$ at $\Lambda$ without excluding spinor kinetic terms that are indispensable to making other fields dynamical at lower energy scales?
We will show that this is naturally achieved by imposing the existence of the degenerate limit of vierbein, $\det e\to0$, on the bare action. It is noteworthy that the degenerate vierbein necessarily arises in the topology-changing configurations including zero eigenmodes of vierbein in the path integral~\cite{Tseytlin:1981ks,Horowitz:1990qb}. 
It would be naturally expected that such topology-changing processes significantly take place around $\Lambda$ where quantum gravitational fluctuations become relevant.

In this Letter, we will show that the finiteness of action in the degenerate limit restricts us to write down only three possible LL invariant terms: the cosmological constant, the linear term in the LL field strength, and the spinor kinetic term, whose coefficients are in general a function of LL singlets such as a scalar field and a spinor bilinear.\footnote{
This theory differs from the spinor gravity in which the Lorentz symmetry is only global~\cite{Hebecker:2003iw}, and also from the spectral action of non-commutative geometry, where all the bosonic fields are required by cancellation of a scale anomaly of a fermionic action~\cite{Andrianov:2010nr}, both in the principle and in the resultant action.
}
In particular, it is forbidden to have a kinetic term for a scalar field and for an ordinary YM gauge field.

For our claim that the LL gauge symmetry is a hidden local symmetry, it is essential to show that the prohibited LL gauge kinetic term $\int F\wedge\star F$ is induced by the spinor loop. 
We demonstrate that the kinetic term of the LL gauge field $\omega$ is induced by the spinor loop below $\Lambda$, and $\omega$ acquires a mass of the order of $\Lambda$ at the same time.
Consequently, the LL gauge symmetry is spontaneously broken from the beginning of the gauge field becoming dynamical. Therefore, it is nothing but a hidden local symmetry, from which gravity is emergent.

Our proposal provides a new insight for quantization of gravity in terms of the spontaneously broken LL gauge theory, similarly to the Nambu-Jona-Lasinio (NJL) model which has served for understanding the spontaneous chiral symmetry breaking and 
effective description of hadron generation in low-energy QCD.

\section{Degenerate gravity at UV cutoff}
We take the vierbein $e^\ba{}_\mu$ and the LL gauge field $\omega^\ba{}_{\bb\mu}$ as fundamental DOF to describe gravity at a certain UV cutoff scale $\Lambda$.\footnote{
If one expects further UV completion above $\Lambda$ such as in string theory, our claim is that the effective action at $\Lambda$ becomes Eq.~\eqref{starting action} after integrating out possible (stringy) modes.
}
Here and hereafter, the bold roman letters $\ba,\bb,\dots$ and the Greek ones $\mu,\nu,\dots$ denote the tangent space basis and the spacetime coordinates in a given chart, respectively. Metric field is defined as a composite of vierbein: $g_{\mu\nu}=\eta_{\ba\bb}e^\ba{}_\mu e^\bb{}_\nu$, where $\eta=\diag\fn{-1,1,1,1}$ is the tangent-space metric.

To define the bare action, we make the following two assumptions: (I) The action at $\Lambda$ is invariant under the diffeomorphisms (\textsf{diff}) and the LL transformation based on $SO(1,3)$: 
\al{
&\tx{\textsf{diff}:~}
\begin{cases}
e^\ba{}_\mu\fn{x}\to e^{\pr\ba}{}_\mu\fn{x'}=e^\ba{}_\nu\fn{x}{\p x^\nu\ov\p x^{\pr\mu}}, \\
\omega_\mu\fn{x}\to \omega'_\mu\fn{x'}=\omega_\nu\fn{x}{\p x^\nu\ov\p x^{\pr\mu}},\\
%\omega^\ba{}_{\bb\mu}\fn{x}\to \omega^{\pr\ba}{}_{\bb\mu}\fn{x'}=\omega^\ba{}_{\bb\nu}\fn{x}{\p x^\nu\ov\p x^{\pr\mu}},\\
\psi\fn{x}\to\psi\fn{x},
\end{cases} 
\nn
\label{diff and LL transformations}\\[-6.5ex]
\nn
&\tx{LL:}~
\begin{cases}
e^\ba{}_\mu\fn{x}\to L^\ba{}_\bb\fn{x}e^\bb{}_\mu\fn{x}, \\
\omega_\mu\fn{x}\to \Paren{\p_\mu L\fn{x}} L^{-1}\fn{x}+ L\fn{x}\omega_\mu\fn{x} L^{-1}\fn{x},\\
\psi\fn{x}\to S\Fn{L\fn{x}}\psi\fn{x},
\end{cases}
\nonumber
}
where $S\Fn{L\fn{x}}$ is the LL transformation, $S\fn{L\fn{x}}=1+{1\ov2}\sigma^{\ba\bb}\theta_{\ba\bb}\fn{x}$ for an infinitesimal $L^\ba{}_\bb\fn{x}=\delta^\ba_\bb+\theta^\ba{}_\bb\fn{x}$ with $\sigma^{\ba\bb}=\commutator{\gamma^\ba}{\gamma^\bb}/4$ being the LL generators, and we employ the short-hand notation for the LL gauge field: $\pa{\omega_\mu}^\ba{}_\bb=\omega^\ba{}_{\bb\mu}$, $\pa{ L\omega_\mu L^{-1}}^\ba{}_\bb= L^\ba{}_\bc\omega^\bc{}_{\bd\mu}\pa{ L^{-1}}^\bd{}_\bb$, etc. 

(II) The bare action at $\Lambda$ must admit any degenerate limit $\ab{e}:=\det_{\ba,\mu}e^\ba{}_\mu\to0$.
The requirement for the action to be finite in the degenerate limit guarantees that we may freely take one or some of the eigenvalues of the vierbein to be zero.
As said above, the degenerate vierbein necessarily arises in the topology-changing configurations~\cite{Tseytlin:1981ks,Horowitz:1990qb} in the path integral, and it would be reasonable to require the existence of the degenerate limit.

The vierbein belongs to the fundamental representation of the LL symmetry and its vacuum expectation value (VEV) spontaneously breaks the LL symmetry.
This is in parallel to the Higgs field being the fundamental representation of the Standard-Model gauge symmetry and its VEV breaks the gauge symmetry. 
In effect, assumption (II) forbids inverse of vierbein $e^{-1}\sim e_\ba{}^\mu$ in the bare action.
This is again in parallel to the fact that we do not introduce the inverse of the Higgs-field-squared $\paren{H^\dagger H}^{-1}$ in the Standard Model, assuming the existence of the weak field limit $H\to0$, even though $\paren{H^\dagger H}^{-1}$ is not forbidden by any symmetry.\footnote{
There is no consensus on the space of gravitational configurations that the path integral is supposed to integrate over; at least in a Euclidean path integral, the action of degenerate configurations is typically infinite so that their contribution to the path integral vanishes (barring the issue of unboundedness of Euclideanization of gravitational system coupled with matter; see e.g.\ Ref.~\cite{Vilenkin:2002ev} for a review). This situation is in parallel to the Higgs analogy: The contribution from $H^\dagger H\to0$ in the action $S_\tx{E}={\kappa/H^\dagger H}+\cdots$ to the path integral vanishes if $\kappa>0$. Such a term is dropped by hand by the assumption of existence of $H\to0$ limit, even though it is superrenormalizable and more relevant than any other ordinary term.
}
So to say, the degenerate limit ensures a ``linear realization" of the LL symmetry.\footnote{
Formally, transformations of the inverse of vierbein are also linear: $e_\ba{}^\mu\fn{x}\to L_\ba{}^\bb\fn{x} e_\bb{}^\mu\fn{x}$ and $e_\ba{}^\mu\fn{x}\to \paren{\p x^{\pr\mu}/\p x^\nu}e_\ba{}^\nu\fn{x}$. Therefore precisely speaking, the terminology ``linear realization'' should rather be understood as the assumption itself, namely, only the vierbein $e^\ba{}_\mu$ is the fundamental degree of freedom and the action does not contain its inverse.
}

Under two assumptions (I) and (II), we find that only the following three terms are compatible with the ``linear realization'' and are relevant for quantum dynamics at $\Lambda$:
\al{
S_\B
	&=	\int\df^4x\ab{e}\bigg[
			-V_\B+{M_\B^2\ov2}e_\ba{}^\mu e_\bb{}^\nu F^{\ba\bb}{}_{\mu\nu} \nn
		&\phantom{\int\df^4x\ab{e}\bigg[-V}	-{Z_\B\ov2}\paren{\overline\psi e_\ba{}^\mu
			\gamma^\ba\mc D_\mu\psi+\tx{h.c.}}
			\bigg],
			\label{starting action}
}
where
$F^\ba{}_{\bb\mu\nu}=\pa{\p_\mu\omega_\nu-\p_\nu\omega_\mu+\commutator{\omega_\mu}{\omega_\nu}}^\ba{}_\bb$ is the LL field strength and
$\mc D_\mu=\p_\mu+{1\ov2}\omega_{\ba\bb\mu}\sigma^{\ba\bb}+iA_\mu^aT^a$ is the covariant derivative associated with the LL and ordinary gauge symmetries, with $T^a$ being the generators of YM gauge transformation.
Here $V_\B$, $M_\B^2$, and $Z_\B$ are the potential, the Planck mass-squared parameter, and the spinor field renormalization factor, respectively, which are in general arbitrary functions of singlets under both \diff and the LL transformation such as $\phi$, $\overline\psi\psi$, etc.
We see that the linear realization severely restricts the possible form of terms at $\Lambda$.

The following comments are in order:
\begin{enumerate}[(i)] 
\item In the ordinary YM gauge theory, the internal gauge space and the spacetime are independently defined and never mix with each other.
In our transformation law~\eqref{diff and LL transformations}, we have separated the LL gauge symmetry and the spacetime \textsf{diff} as in the ordinary YM theory.
However, once vierbein acquires a non-zero VEV $\bar e^\ba{}_\mu$  at lower energies, these two spaces necessarily mix with each other.
This is a distinct aspect of the LL gauge theory from the YM one, in addition to being able to write down the gauge invariant linear action.
This is an essential point in understanding why the spontaneous LL-symmetry breaking plays a crucial role of the generation of spacetime. 

\item The apparent existence of the inverse vierbein $e_\ba{}^\mu$ in the action~\eqref{starting action} is spurious since it disappears when combined with the determinant: $\ab{e}e_\ba{}^\mu={1\ov3!}\ep{\ba\bb\bc\bd}\ep{\mu\nu\rho\sigma}e^\bb{}_\nu e^\bc{}_\rho e^\bd{}_\sigma$ and $\ab{e}e_{[\ba}{}^\mu e_{\bb]}{}^\nu={1\ov2!2!}\ep{\ba\bb\bc\bd}\ep{\mu\nu\rho\sigma}e^\bc{}_\rho e^\bd{}_\sigma$, where the summation with the totally anti-symmetric symbol with $\ep{0123}=1$ is understood.

\item The ``linear realization'' of the LL symmetry forbids the scalar and ordinary gauge kinetic terms $-{1\ov2}\ab{e} g^{\mu\nu}\tr(\paren{D_\mu\phi}^\dagger D_\nu\phi)$ and $-{1\ov2g^2}\ab{e}g^{\mu\rho}g^{\nu\sigma}\tr(G_{\mu\nu}G_{\rho\sigma})$, where $g$ in the denominator is a gauge coupling, $D_\mu=\p_\mu+iA_\mu$.\footnote{
Suppose that we take the zero field limit, $e^\ba{}_\mu\to0$, for all the components uniformly. Then, since we have four vierbein components from $\ab{e}$, they are enough to cancel two inverse vierbein from $g^{\mu\nu}$. In contrast, when we take the degenerate limit in which we only make part of eigenvalues of $e^\ba{}_\mu$ to be zero such that there is some components of inverse vierbein that do not go to zero, there remains some component of $g^{\mu\nu}$ that is left constant and hence $\ab{e}g^{\mu\nu}$ diverges in general. In this sense, the degenerate limit is wider than the uniform zero limit as the former includes the latter as a part of it.
}

\item It is also forbidden to put the Levi-Civita connection $\Gamma^\mu{}_{\rho\sigma}={g^{\mu\nu}\ov2}\paren{-\p_\nu g_{\rho\sigma}+\p_\rho g_{\sigma\nu}+\p_\sigma g_{\nu\rho}}$ and the Levi-Civita spin connection $\Omega^\ba{}_{\bb\mu}=e^\ba{}_\lambda\paren{ \p_\mu e_\bb{}^\lambda+\Gamma^\lambda{}_{\sigma\mu}e_\bb{}^\sigma}$.

\item In principle, we can also add the so-called Euler, Pontryagin, Nieh-Yan, and Immirzi terms, which do not involve the inverse of vierbein; see Ref.~\cite{Freidel:2005sn}. These terms are topological or exact, and hence we omit them here for simplicity; see also Ref.~\cite{Daum:2013fu} for discussion on the special case of (anti) self-dual choice of the Immirzi parameter.
\end{enumerate}

At this level of the action \eqref{starting action}, there are no apparent kinetic terms for both $e$ and $\omega$, while there is a mixing term such as $ee\p\omega$ as well as their interaction term with the spinor field.
One may regard $e$ and $\omega$ as auxiliary fields. 
Their kinetic terms will be generated at the loop level 
by the spinor field fluctuations, 
as we will see later. 
In this sense, these auxiliary fields might be interpreted as composite fields of spinor fields, which will become dynamical below $\Lambda$ at the loop level.  
We may also recall the compositeness condition~\cite{Bardeen:1989ds}, with which the action \eqref{starting action} at the cutoff scale $\Lambda$ is a boundary condition of the low energy effective theory for this system. 

Let us clarify our stance in proposing the action~\eqref{starting action} in analogy with the NJL model; see Ref.~\cite{Hatsuda:1994pi} for a review. The NJL model has QCD as its UV completion, and the linear $\sigma$/quark-meson models as its infrared (IR) effective field theory.\footnote{
More precisely, the linear $\sigma$ model is obtained by integrating out fermions from the quark-meson model and discarding the perturbatively non-renormalizable terms. Here we do not distinguish them in this analogy; see also the discussion below.
}
As we lower the energy scale further, we end up with the non-linear $\sigma$ model.
The theory~\eqref{starting action} is analogous to the NJL model in the sense that it has general relativity (GR) as its low energy effective theory, and is supposed to have (yet unknown) UV completion above; see the table below.
\begin{widetext}
\begin{center}
\begin{tabular}{c||c|c}
%\hline
& Strong interaction & Gravity \\
\hline
$E>\Lambda$ & 
	QCD & 
	A conceivable UV completion\\
$E=\Lambda$
	& NJL model at $\Lambda\sim\Lambda_\tx{QCD}$
	& Theory~\eqref{starting action} at $\Lambda\sim\MP$\\
$E<\Lambda$
	& Linear $\sigma$/quark-meson model with dynamical $\rho$
	& Effective theory of~\eqref{starting action} with dynamical $\omega$ \\
$E\ll\Lambda$
	& Non-linear $\sigma$ model
	& GR without $\omega$ \\
%\hline
\end{tabular}
\end{center}
\end{widetext}
More precisely, we will see that the theory~\eqref{starting action} dynamically generates the kinetic term for the LL gauge field~$\omega$ below $\Lambda$ from a spinor loop. This corresponds to the dynamical generation of the $\rho$ meson field (hidden-local gauge field) by integrating out the higher frequency modes of fermions in the NJL model, after which the resultant effective theory becomes a quark-meson model including the $\rho$ meson field, analogously to the theory below $\Lambda$ that have dynamical $\omega$.

In the linear $\sigma$/quark-meson model as a low energy effective theory of the NJL model, the field renormalization factors in the kinetic terms of all the hadronic fields go to zero as we raise the energy toward $\Lambda_\tx{QCD}$ from below in the renormalization group flow. 
The forbidden kinetic terms in the action~\eqref{starting action}, due to the requirement of existence of degenerate limit, would correspond to this vanishing kinetic term in the quark-meson model.
Further pushing this analogy, one might interpret the vierbein and/or LL gauge field as a composite of some spinor fields.
To establish this analogy, one needs to verify that a pole corresponding to the composite field appears in the scattering amplitude of the constituent spinors.
Then the composite field can indeed be regarded as an auxiliary field written by the constituent spinors through an equation of motion, analogously to the bosonization of the NJL model.
At the moment, we leave it open whether or not the LL gauge field and/or vierbein (or even the Higgs and ordinary gauge fields) become composite in a UV completion of our model.%\footnote{
%Even if the vierbein and LL gauge field are indeed composite, the vanishing kinetic terms for the Higgs and ordinary gauge fields do not necessarily indicate that they are composite too: It is also possible that SM Higgs and gauge bosons are not composite, like photons and leptons for the strong dynamics.
%}

\section{Generation of LL gauge kinetic term}
We demonstrate that a spinor loop generates a kinetic term for the LL gauge field, starting from the (bare) action~\eqref{starting action} with only taking into account the spinor mass term $m\overline\psi\psi$ in~$V_\B$ and with regarding $M_\B^2$ and $Z_\B$ as constants, for simplicity. When $Z_\B$ is constant, we may redefine the spinor field such that $Z_\B=1$, as we will do hereafter.

The vierbein background $\bar e$ should be determined dynamically by a stationary condition for the quantum-dressed effective potential $V_\tx{eff}$ at low energy.
A possible approach to this issue is to first make an ansatz for $\bar e$, compute $V_\tx{eff}$ that depends on $\bar e$, and examine the self-consistency condition from its stationary condition; see Ref.~\cite{Floreanini:1991cw}.

Our approach in this paper is similar to the treatment in electroweak theory before the discovery of Higgs particle: Although the mechanism of electroweak symmetry breaking was not established, one had set $\langle H^\dagger H \rangle=v_h^2/2$ with $v_h=246$\,GeV, and had computed predictions on that assumption.
In gravitational theory, a flat background field is a simple solution to the Einstein theory with no cosmological constant.
Therefore, it is reasonable to discuss the dynamics of gravity by choosing a specific background field as a first step.

In this Letter, we start from an ansatz of a flat spacetime background $\bar e^\ba{}_\mu=C\delta^\ba_\mu$ and $\bar\omega^\ba{}_{\bb\mu}=0$, where $C\to0$ corresponds to a degenerate limit in the symmetric phase, whereas $C\neq0$ to the broken Higgs phase; see Discussion below.
Here, we concentrate on the theoretical motivation whether or not it is valid that we claim the LL gauge symmetry is a hidden local symmetry.
To this end, we concentrate on generation of the LL gauge kinetic term in this work, and leave the evaluation of $V_\tx{eff}$ for future study.\footnote{
We do not discuss the kinetic-term generation for the scalars and the ordinary gauge bosons, which can be trivially done as in the NJL model describing chiral symmetry breaking in QCD; see e.g.\ Ref.~\cite{Hatsuda:1994pi} for a review.
}
When there is not (yet) the kinetic term for the vierbein as in the bare action~\eqref{starting action}, we may always redefine the vierbein field for $C\neq0$. Hereafter we put an ansatz $C\neq0$ and set $C=1$.

We calculate the kinetic term for the LL gauge field ${Z_\omega\ov2} g^{\mu\nu} g^{\rho\sigma} F^\ba{}_{\bb\mu\rho}F^\bb{}_{\ba\nu\sigma}$, which contains a term
\al{
-{Z_\omega\ov2} \omega_{\ba\bb\mu}\paren{\eta^{\bc[\ba}\eta^{\bb]\bd}\paren{\eta^{\mu\nu}\p^2-\p^\mu\p^\nu}}\omega_{\bc\bd\nu},
		\label{Kinetic term for LL gauge field}
}
where $Z_\omega$ is the field-renormalization factor and the square brackets for indices denote anti-symmetrization.
Generation of a finite value of $Z_\omega$ indicates that the LL gauge field has become dynamical.
The kinetic operator \eqref{Kinetic term for LL gauge field} is induced from the two-point function of the LL gauge field:
\vspace{-0.4cm}
\als{
I^{\ba\bb\bc\bd\mu\nu}\fn{p}
	&=	\vcenter{\hbox{\includegraphics[width=40mm]{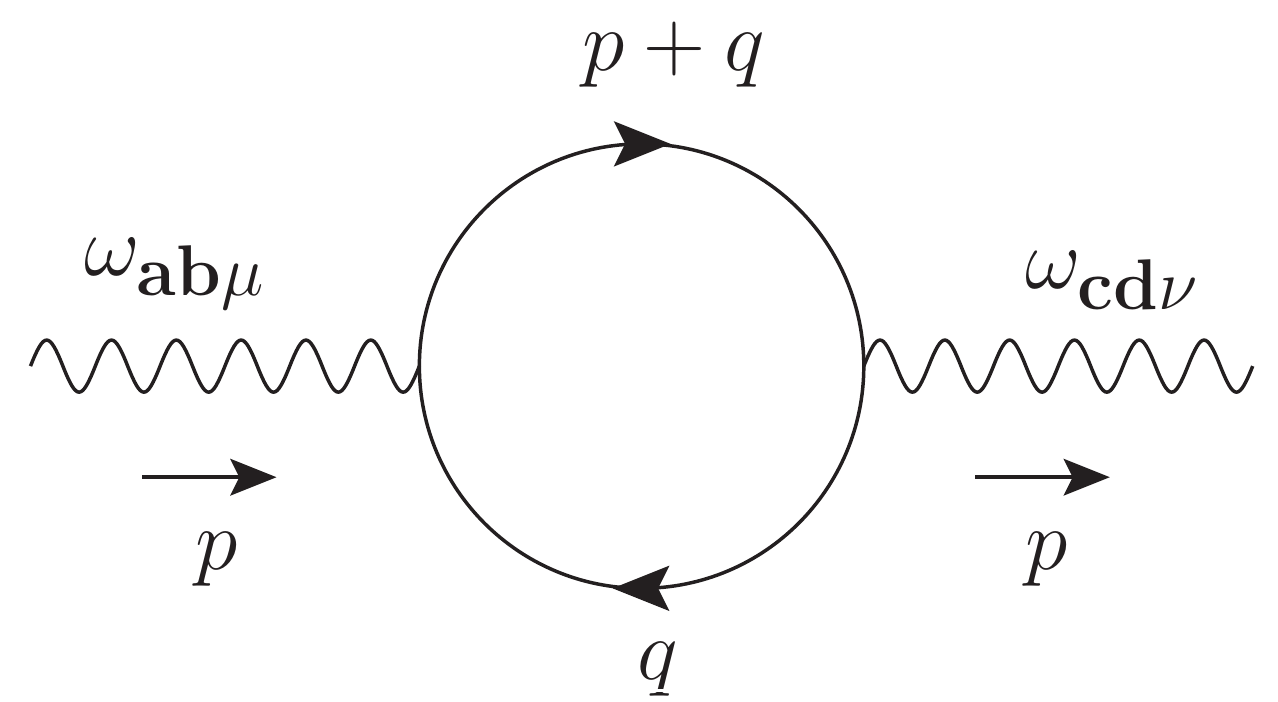}}\vskip0mm}.
}

After some computation, we obtain a result containing the tensor structure corresponding to Eq.~\eqref{Kinetic term for LL gauge field},
$
I^{\ba\bb\bc\bd\mu\nu}\fn{p}
	\supset
		-i\eta^{\bc[\ba}\eta^{\bb]\bd}\sqbr{
			f\fn{p^2}\paren{\eta^{\mu\nu}-{p^\mu p^\nu\ov p^2}}
			+g\fn{p^2}{p^\mu p^\nu\ov p^2}
			},
$ 
with the form factors
\al{
f\fn{p^2}
	&=	m_\omega^2+{1\ov128\pi^2}{1\ov\paren{1+{m^2\ov\Lambda^2}}^2}p^2+\cdots,
\nn
\label{f and g}\\[-6ex]
\nn
g\fn{p^2}
	&=	m_\omega^2-{1\ov96\pi^2}\paren{{\ln{\Lambda^2+m^2\ov m^2}-{1\ov1+{m^2\ov\Lambda^2}}}}p^2+\cdots,\nonumber
}
where we have cut off the momentum integral by $\Lambda$;
the dots denote higher powers of ${p^2\ov\Lambda^2+m^2}$; 
and 
$m_\omega^2={\Lambda^2\ov64\pi^2}\paren{{\Lambda^2+4m^2\ov\Lambda^2+m^2}-{4m^2\ov\Lambda^2}\ln{\Lambda^2+m^2\ov m^2}}$ is a quadratically divergent mass-renormalization constant.

Now we can read off~$Z_\omega=\df f(p^2)/\df p^2|_{p^2=0}$ for ${m^2/\Lambda^2}\ll1$: $Z_\omega={1/(128\pi^2)}$.
It is remarkable that both the logarithmic and quadratic divergences have canceled out in~$Z_\omega$.
As we define the renormalized field $\omega_\tx{R}:=\sqrt{Z_\omega}\omega$ to canonically normalize its kinetic term,
its mass $M_\omega$ becomes of the order of $\Lambda$: $M_\omega = m_\omega/\sqrt{Z_\omega} \simeq \sqrt{2} \Lambda$ for $m/\Lambda\ll 1$. 
To conclude, we have found that the LL gauge field acquires the kinetic term and becomes a~dynamical field.\footnote{
For reader's reference, we show the full form:
\als{
&I^{\ba\bb\bc\bd\mu\nu}\fn{q}=
{\widehat I}_{\rho\sigma}\fn{q}\, e_{\ba'}{}^{\mu} e_{\bb'}{}^{\rho} e_{\bc'}{}^{\nu} e_{\bd'}{}^{\sigma}\nn
&\,\times\paren{\epsilon^{\ba\bb\ba'\bd'}\epsilon^{\bc\bd\bc'\bb'} + \epsilon^{\ba\bb\ba'\bb'}\epsilon^{\bc\bd\bc'\bd'} 
 - \epsilon^{\ba\bb\ba'\be}\epsilon^{\bc\bd\bc'\be'}\eta_{\be\be'}\eta^{\bb'\bd'}
 	}\nn
&+
 {\widehat J}\fn{q}\, e_{\ba'}{}^{\mu}  e_{\bc'}{}^{\nu}\epsilon^{\ba\bb\ba'\be}\epsilon^{\bc\bd\bc'\be'}\eta_{\be\be'},
}
where $\widehat I_{\rho\sigma}\fn{q}$ and $\widehat J\fn{q}$ are the standard loop functions:
\als{
{\widehat I}_{\rho\sigma}\fn{q}
&=\int{\df^4p\ov\paren{2\pi}^4} \frac{(p+q)_{\rho} p_{\sigma}}{[(p+q)^2-m^2][p^2-m^2]},\\
{\widehat J}\fn{q}&=\int{\df^4p\ov\paren{2\pi}^4} \frac{m^2}{[(p+q)^2-m^2][p^2-m^2]}.
}
This contains the contributions to the other terms such as the ``Ricci-tensor-squared'' $F_{\ba\mu}F^{\ba\mu}$ and ``Ricci-scalar-squared'' $F^2$, where $F_{\ba\mu}:=e^{\bb\nu}F_{\ba\bb\mu\nu}$ and $F:=e^{\ba\mu}F_{\ba\mu}$, as well as the topological terms such as the Pontryagin term $F^{\ba\bb}\wedge F_{\ba\bb}$ and the Euler term $\epsilon^{\ba\bb\bc\bd}F_{\ba\bb}\wedge F_{\bc\bd}$. In this Letter we focus on the validity of our claim that the LL gauge symmetry is a hidden local symmetry and have only shown the coefficients~\eqref{f and g}. From the above expressions, we see that our results do not change if we employ the dimensional regularization instead of the naive cutoff.
}

\newpage
Naively one might find it obvious that a charged spinor gives a contribution to a YM kinetic term via loop diagrams, following from standard heat kernel formulas. However, the heat kernel method can only take into account \emph{divergent} contributions. Indeed, there are earlier works based on the heat-kernel expansion, and it is concluded that the field renormalization of LL gauge fields is \emph{not} generated by spinor loop~\cite{Tseytlin:1981ks}. We, for the first time, have shown the generation of kinetic term for the LL gauge field by directly computing the \emph{finite} correction.

We have also computed the same vacuum polarization diagram for the vierbein.
We have found that the trace mode of (dimensionless) vierbein acquires a quartically divergent mass operator around the symmetric phase $\bar e^\ba{}_\mu=0$, 
as well as a logarithmically divergent kinetic term.

\section{Discussion}

The LL gauge symmetry is spontaneously broken once the vierbein background is determined to be any non-zero value such as the flat spacetime  in the above example,
whereas the zero-field limit $\bar e^\ba{}_\mu\to0$ corresponds to the symmetric phase in the Higgs mechanism.
The vierbein field plays the role of the Higgs field, being a fundamental representation of the LL gauge symmetry. A difference here is that the ``Higgs'' field is a vector field, unlike in an ordinary Higgs mechanism in which it is a scalar.
Once vierbein kinetic terms are generated from loop effects, its LL covariant derivative in the broken phase $\bar e^\ba{}_\mu\neq0$ will also contribute to the mass of the LL gauge field through the Higgs mechanism, to be added to $m_\omega^2$ above.
Indeed, our ansatz $\bar e^\ba{}_\mu=C\delta^\ba_\mu$ with a finite constant $C$ corresponds to the Higgs field obtaining an expectation value $\bar H^a =v_H\delta^{a1}\neq 0$ in the Standard Model as a solution of the stationary (self-consistency) condition, leading to the massive gauge bosons $W$ and $Z$.

The vierbein background induces, via the spinor loop, not only the transverse term with $f\fn{p^2}$ but also directly the non-vanishing longitudinal term with $g\fn{p^2}$ in Eq.~\eqref{f and g}.
This is the very characteristic of the LL gauge symmetry,
in contrast to the usual Higgs mechanism, where the gauge symmetry is only broken by the mass term of the gauge field. 

Our result is consistent with the Weinberg-Witten theorem~\cite{Weinberg:1980kq}:
The above-mentioned longitudinal term results in the violation of the naive current conservation, $p_\mu I^{\ba\bb\bc\bd\mu\nu}\fn{p}\neq0$.
That is, the LL-current conservation is spontaneously broken as soon as the LL gauge field becomes dynamical no matter whether the LL gauge field becomes massive or not.
This implies that the LL gauge symmetry is necessarily spontaneously broken, that is, the LL symmetry is nothing but the hidden local symmetry~\cite{Bando:1987br}.
In other words, the theory at $\Lambda$ retains the LL gauge invariance even when we integrate out the auxiliary LL gauge field, where the LL gauge invariance is ``hidden'' in the UV scale physics and is dynamically emergent in the infrared-scale physics, analogously to the hidden local symmetry carried by $\rho$ meson in the low-energy QCD.

In the original action at $\Lambda$, there is a local $\diff\times SO(1,3)$ symmetry, which has $4+6=10$ DOF.
Within 16 DOF of vierbein fluctuations, the 6 modes of the $SO(1,3)$ Nambu-Goldstone direction are eaten as the longitudinal modes of the LL gauge field, while the remains correspond to the 10 classical DOF of the graviton.
Among them, 4 modes of vierbein fluctuations are reduced by the transverse condition for the \textsf{diff}. At the quantum level, remaining 6 DOF are further subtracted by the loop of 4 \diff ghost fields, resulting in the 2 DOF of quantum graviton fluctuation.\footnote{This spontaneous breaking pattern is different from that of the induced gravity, in which the gravitons arise as Nambu-Goldstone bosons associated with the breaking structure of $GL(4) \to SO(1,3)$~\cite{Nakanishi:1979fg}.
}

%As we lower the scale below $\Lambda$, the LL gauge field appears to become dynamical at %the lower loop level than the vierbein (except for its trace mode). It may be intriguing that %the LL gauge field seems to become dynamical earlier than the vierbein, which should be %compared to the common classical procedure to integrate out the LL gauge field first as an %auxiliary field to make the vierbein dynamical.

Let us discuss possible future directions in the following paragraphs:

%In this Letter we have taken the action~\eqref{starting action} as a low energy effective %action:
%The fluctuating mode of vierbein is charged under the spontaneously broken (hidden) LL %gauge symmetry, and the situation is analogous to the hidden local symmetry below the QCD %scale.
%One might see it as a hint for a quantum gravity that uses the renormalization procedure of %the hidden local symmetry at the loop level, on the same footing as the chiral perturbation %theory; see e.g.\ Ref.~\cite{Harada:2003jx} for a review.

We have assumed that the flat vierbein background becomes a vacuum solution. On physical ground, we expect that the vierbein fluctuation will become massless around a vacuum $\bar e^\ba{}_\mu\neq0$ in the end. To confirm this expectation, we need to compute the full effective potential.
In this Letter, we have computed the spinor loop correction with the naive momentum cutoff. It is important to improve it by a non-perturbative method such as the functional renormalization group. 
%These points will be pursued in a separate publication.

It is also necessary to verify the vanishing kinetic term in the renormalization-group flow toward UV direction within the effective theory below $\Lambda$.

It is worth studying not only UV but also IR fixed-point structure in the proposed theory (at $E<\Lambda$ in the table above). This theory might have a Caswell-Banks-Zaks-like fixed point in the IR limit since it is a large flavor non-Abelian gauge theory, given the SM spinor degrees of freedom.
Such an IR fixed point, if exists, could belong to the same universality class to which a UV fixed point of asymptotically safe gravity belongs~\cite{Hawking:1979ig,Reuter:1996cp,Souma:1999at,Daum:2013fu}. (Asymptotically safe gravity in this context should correspond to the line $E\ll\Lambda$ in the table above.)
One may remind that in three spacetime dimensions, there is an example of known such equivalence of UV and IR limits of the IR and UV theories, respectively: The UV fixed point of the (IR) non-linear $\sigma$ model and the IR fixed point of the (UV) linear $\sigma$ model belong to the same universality class~\cite{Flore:2012ma}.

In this Letter we have not included possible loop effects that are generated if we take into account the EH term, the second term in Eq.~\eqref{starting action}, as kinetic mixing between the vierbein and the LL gauge field.\footnote{
It is rather hard to define a quantum field theory with a proper cutoff scheme if the Einstein-Cartan term is the only kinetic term among these fields, as is discussed in Sec. 3.2 (i) in Ref.~\cite{Daum:2013fu}. 
}
Instead, we have demonstrated that the kinetic terms for the vierbein and LL gauge field are induced from the spinor loop. 
In the complete treatment of the hidden local symmetry, we should include all the possible induced terms, including the kinetic terms and the EH one, and examine whether the compositeness condition can be satisfied at $\Lambda$, namely, whether all but the ones in Eq.~\eqref{starting action} vanish as we raise the energy scale upwards.

%{\it Acknowledgements}
\subsection*{Acknowledgements}
We thank Hidenori Fukaya, Yuta Hamada, Taishi Katsuragawa, Hikaru Kawai, Taichiro Kugo, Norihiro Iizuka, Robert Percacci, Christof Wetterich, Satoshi Yamaguchi, and especially Kengo Kikuchi for valuable discussions and comments.
K.O.\ thanks the hospitality of Center for Theoretical Physics and College of Physics, Jilin University where the present work has been partially done.
This work was supported in part by the National Science Foundation of China (NSFC) under Grant No.\,11747308 and 11975108, 
and the Seeds Funding of Jilin University (S.M.).
The work of K.O.\ is in part supported by JSPS Kakenhi Grant No.~19H01899.
M.Y.\ is supported by the Alexander von Humboldt Foundation.

\bibliographystyle{JHEP} 

\bibliography{refs}
\end{document}